\documentclass[twocolumn,showpacs,prd,floatfix]{revtex4}
\pdfoutput=1
\usepackage{graphicx}

\newcommand{\beq}{\begin{equation}}
\newcommand{\eeq}{\end{equation}}
\newcommand{\bea}{\begin{eqnarray}}
\newcommand{\eea}{\end{eqnarray}}

\newcommand{\lie}{\mathcal{L}}

\newcommand{\gtt}{g_{\theta\theta}}

\newcommand{\Ktt}{K_{\theta\theta}}
\newcommand{\for}{\quad\mbox{for }}

\newcommand{\M}{}

\usepackage{color}

\begin{document}

\title{
Schwarzschild black hole as moving puncture in isotropic coordinates
}

\author{Bernd Br\"ugmann}

\affiliation{Theoretical Physics Institute, 
University of Jena, 07743 Jena, Germany}

\date{April 28, 2009}

\begin{abstract}
The success of the moving puncture method for the numerical simulation
of black hole systems can be partially explained by the properties of
stationary solutions of the 1+log coordinate condition.  We compute
stationary 1+log slices of the Schwarzschild spacetime in isotropic
coordinates in order to investigate the coordinate singularity that
the numerical methods have to handle at the puncture.  We present an
alternative integration method to obtain isotropic coordinates that
simplifies numerical integration and that gives direct access to a
local expansion in the isotropic radius near the puncture. Numerical
results have shown that certain quantities are well approximated by a
function linear in the isotropic radius near the puncture, while here
we show that in some cases the isotropic radius appears with an
exponent that is close to but unequal to one.
\end{abstract}


\maketitle

\section{Introduction}
\label{introduction}

The moving puncture method \cite{CamLouMar05,BakCenCho05} is the basis
of many of the recent successful simulations of black hole
binaries in numerical relativity. For black hole puncture data, the
term puncture refers to a single point on a hypersurface where the
metric has a coordinate singularity that characterizes the presence of
a black hole, while the physical singularity is not part of the
hypersurface.

The original puncture data~\cite{BraBru97,BeiOMu94,BeiOMu96,DaiFri01} uses the
Brill-Lindquist ``wormhole'' topology \cite{MisWhe57,BriLin63}, where
the hypersurface connects an outer with an inner asymptotically flat
region, and the coordinate singularity occurs when compactifying the
inner end by a singular conformal factor in isotropic coordinates. For
the Schwarzschild solution in isotropic coordinates with radius $r$
the conformal factor is $\psi=1+\frac{M}{2r}$, where $M$ is the mass,
and the puncture singularity occurs for $r\rightarrow0$ and
Schwarzschild areal radius $R\rightarrow\infty$. The moving puncture
method typically uses wormhole data as initial data, however the gauge
conditions (1+log slicing and Gamma-freezing shift) lead after some
gauge evolution to locally stationary solutions that are perhaps
better described by the term
``trumpet''~\cite{HanHusPol06,Bro07a,GarGunHil07,HanHusOhm08}. 
Trumpets for a stationary maximal slice are discussed
in~\cite{HanHusOMu06,BauNac07}.
For Schwarzschild in isotropic coordinates, the stationary 1+log slice
is characterized by $\psi\sim\frac{1}{\sqrt{r}}$ for small $r$, and
the trumpet ends at a finite Schwarschild radius, $R\rightarrow
R_0\approx 1.312M$ for $r\rightarrow0$.

The analytic stationary 1+log solution was first described for moving
punctures in \cite{HanHusPol06} (see \cite{BucBar05} for related
results that however do not address the issue of black hole
punctures). The analytic solution has been used to develop quite a
complete picture of the geometry and global properties of the
Schwarzschild solution for stationary 1+log slices
\cite{Bro07a,GarGunHil07,HanHusOhm08,Ohm07}, and detailed information about
coordinate effects near the puncture has also been obtained through
the analysis of 1D numerical simulations in spherical symmetry
\cite{Bro07a,GarGunHil07,Bro07}. 

In \cite{HanHusPol06}, we also discussed regularity at the puncture in
terms of a Taylor expansion in the isotropic coordinate radius $r$
around $r=0$, similar in spirit to previous work on fixed
punctures~\cite{AlcBruDie02,Bru97}. The motivation for such an investigation
is to find out what sort of singularity a typical 3D BSSN code in
Cartesian coordinates has to handle at the puncture. On the one hand,
the analytic results have shown that a solution for 1+log slicing
exists, and it is an experimental fact that numerical codes are able
to approximate this solution. On the other hand, it is not yet fully
understood how the finite difference codes succeed in obtaining
accurate approximations to a black hole puncture that evidently is not
regular at the puncture. For example, in the numerical coordinates the
conformal factor and the lapse are not smooth at the puncture, and the
BSSN extrinsic curvature displays a finite jump
discontinuity~\cite{HanHusPol06}.
The analysis in~\cite{HanHusPol06} suggested that at least some of the
leading order behavior near $r=0$ could be reliably obtained by a
Taylor expansion since it approximates the numerical results rather
well, although the expansion was at least partially an ad hoc
ansatz. The existence of a solution with a particular small $r$
behavior was assumed, and its consistency and some consequences
were derived by inserting the ansatz into the full BSSN system and the
gauge conditions.

The purpose of the present paper is to derive the small $r$ series in
isotropic coordinates from the analytic solution for stationary 1+log
slices of the Schwarz\-schild solution. The focus is on the calculations
since as mentioned above the geometric picture has already been
discussed elsewhere.
We assume that isotropic coordinates approximate the 3D numerical
simulations well near the puncture, although in fact the 3D numerical
coordinates are not isotropic. We leave the comparison to actual
simulations, as well as the question how finite differencing works in
this case to future work. The key question addressed here is what the
analytic results imply for the singularities at the puncture in
isotropic coordinates.

The main result is that the leading order behavior in some quantities
is indeed described by terms linear in $r$ as assumed in the previous
ansatz.  In particular, imposing stationarity of the conformal factor
in the given coordinates, $\partial_t\psi = 0$, led to the conclusion
that due to the combined evolution and gauge equations
$\psi\simeq\sqrt{R_0/r}$ for small $r$, where $R_0$ is the
Schwarzschild areal radius at the puncture point. 
Furthermore, the lapse vanishes at the puncture, $\alpha\simeq0$, and
the shift vector satisfies $\beta^i\sim r n^i$, where
$n^i=x^i/r$. These results also hold in our new analysis.  However,
the analysis of the present paper shows that non-leading order terms
involve factors of $r$ to some non-integer power.  Making the point,
the previous ansatz was
\beq
	\psi^{-2}\sim r, \quad \beta^r\sim r, \quad \alpha\sim r,
\eeq
while as we show here the analytic stationary 1+log slice
implies for isotropic coordinates
\beq
	\psi^{-2}\sim r, \quad \beta^r\sim r, \quad \alpha\sim r^{1.091}.
\label{leadord}
\eeq
As a second result, the methods we develop for the integration of the
isotropy condition are valid for all $r$ (not just locally near $r=0$),
and they allow a somewhat simpler way to compute initial data for a
Schwarzschild puncture in the moving puncture gauge than was available
previously~\cite{HanHusOhm08}.  

In Sec.~\ref{equations}, we establish our notation for the
Schwarzschild solution in 3+1 form. In Sec.~\ref{onepluslog}, we
summarize results for stationary 1+log slicing of the Schwarzschild
solution. Sec.~\ref{isotropic} describes a novel way to perform
various integrations required to obtain isotropic coordinates. In
Sec.~\ref{numerics} we comment on numerical methods to compute the
1+log data. Sec.~\ref{discussion} concludes with a discussion.

We use geometric units $G=c=1$, and we set $M = 1$ in calculations,
although for clarity we give factors of $M$ in some places.


\section{Schwarzschild solution in 3+1 form}
\label{equations}

The basic equations for the Schwarzschild solution in a 3+1
decomposition 
can be written in a number of ways. We choose to start with
coordinates that encompass both Schwarzschild coordinates, which are
convenient for solving the Einstein equations, and spatially isotropic
coordinates, which are typically used for puncture initial data.


\subsection{Time-independent, spherically symmetric metric}

Consider the metric in 3+1 form assuming spherical symmetry and time
independence. We introduce coordinates $t$, $r$, $\theta$, and $\phi$
for which the metric takes the form
\bea
  \mbox{}^{(4)}ds^2 &=& - (\alpha^2 - \beta^2) dt^2 + \frac{2\beta}{f} dt dr
\nonumber \\
      &&+ \frac{1}{f^2} dr^2 + R^2 (d\theta^2 + \sin^2\theta d\phi^2),
\label{fourds2}
\eea
where the coefficients only depend on $r$. 
The Arnowitt-Deser-Misner (ADM) variables are the 3-metric $g_{ij}$
and the extrinsic curvature $K_{ij}$, plus the lapse $\alpha$ and
shift $\beta^i$, e.g.\ \cite{Yor79}. For
our choice of coordinates
$\alpha = \alpha(r)$,
$\beta^r = \beta(r) f(r)$,
$g_{rr} = 1/f(r)^2$,
$\gtt = R(r)^2$,
$K_{rr} = k(r)/f(r)^2$, and
$\Ktt = l(r) R^2(r)$. Furthermore,
$g_{\phi\phi} = \gtt \sin^2\theta$ and 
$K_{\phi\phi} = \Ktt \sin^2\theta$, and all other components vanish.
The function $\beta$ is the norm of $\beta^i$, 
$
	\beta^2 = \beta^r\beta^r/f^2 = g_{rr}\beta^r\beta^r =
	          \beta_i\beta^i.
$
With $n^i$ the normal vector to constant $r$
surfaces normalized such that $g_{ij} n^in^j=1$, the extrinsic
curvature is expressed using two functions $k(r)$ and $l(r)$,
	$K_{ij} = k n_i n_j + l (g_{ij} - n_i n_j)$,
with trace
\beq
	K = k + 2l.
\label{admK}
\eeq 	

In Schwarzschild radial gauge, the areal radius equals the coordinate radius,
\beq
	R(r) = r,
\eeq
which for vanishing shift, $\beta = 0$, leads to Schwarzschild
coordinates. Isotropic coordinates are given by a conformal factor
$\psi(r)$ such that
\beq
\frac{1}{f^2} dr^2 + R^2 d\Omega^2 = \psi^4 (dr^2 + r^2 d\Omega^2), 	
\eeq
where $dr^2 + r^2 d\Omega^2$ is the flat metric in spherical
coordinates. Isotropic coordinates are therefore given by
$f$ and $R$ that satisfy
\beq
	f = \frac{r}{R}, 
\label{fofrR}
\eeq
and the conformal factor is
\beq
	\psi = \sqrt{\frac{R}{r}}, \quad \psi^{-2} = \frac{r}{R}.
\eeq

\subsection{Solution of the Einstein equations}

The Schwarzschild solution for our choice of coordinates can be
written as
\bea
\alpha &=& f R',
\label{solalpha}
\label{sa}
\\
\beta  &=& \sqrt{\alpha^2 - 1 + \frac{2M}{R}},
\label{solbeta}
\\
k      &=& \frac{\beta'}{R'},
\label{solk}
\\
l      &=& \frac{\beta}{R}.
\label{soll}
\label{sb}
\eea
These relations can be obtained by coordinate transformation from the
standard form of the Schwarzschild solution, but it is also
instructive to derive them by solving the ADM equations starting with
(\ref{fourds2}).  
Given the two metric coefficients $f(r)$ and $R(r)$,
the four remaining quantities for lapse, shift, and extrinsic
curvature are determined. In fact, only $\alpha$ depends directly on the choice
of $f$, and only $\alpha$ involves the coordinate $r$ since the
remaining equations can be expressed in terms of $R$ with $\beta'/R' =
d\beta(R)/dR$. Note that
\beq
\alpha^2 - \beta^2 = 1 - \frac{2M}{R},
\label{a2b2}
\eeq
which shows that we are using the Killing lapse and shift.

As already mentioned, we obtain Schwarzschild
coordinates for $R(r)=r$ and $\beta=0$, which implies
$\alpha^2=1-\frac{2M}{R}$, $f=\alpha$, and $k=l=0$.
Here we are looking for stationary solutions for 1+log slicing with
$\alpha \geq 0$, which is given in terms of one equation that fixes
the lapse, and which leads to a shift and extrinsic curvature which
are not identically zero. We fix the remaining freedom by requiring
either the Schwarzschild radial gauge, $R=r$, or the isotropic gauge,
$f=r/R$.


\section{Stationary 1+log slicing of the Schwarzschild solution}
\label{onepluslog}

We summarize several results on the integration of the stationary 1+log
condition~\cite{HanHusPol06,GarGunHil07,HanHusOhm08,Ohm07,BauEtiLiu08}
while adding some details useful for our purposes, for example a brief
derivation of the integrated stationary 1+log condition in the more
general coordinates of Sec.~\ref{equations}.

\subsection{Integration of the stationary 1+log condition}

The 1+log slicing condition \cite{BonMasSei94} is 
\beq
  (\partial_t - \lie_\beta) \alpha = - 2 \alpha K.
\eeq
If the shift term is absent, which has also been used in some numerical
simulations, then the stationary slice is a maximal slice, $K=0$,
which we will not discuss here (see \cite{HanHusOMu06,BauNac07}).
Manifest stationarity $\partial_t\alpha=0$ implies
\beq
\lie_\beta \alpha = 2 \alpha K,
\eeq
which in our coordinates becomes
\beq
f \beta \alpha' = 2 (k + 2 l) \alpha .
\label{opluslog}
\eeq
This equation looks rather harmless, but for an unfortunate choice of
variables, say when written for the conformal factor in isotropic
coordinates, it becomes a second order ODE in the metric coefficients
that cannot be solved explicitly.
However, an easy way to proceed is to eliminate $f$, $k$, and $l$ (but
not $\beta$) from (\ref{opluslog}) using the Schwarzschild solution
(\ref{sa}) - (\ref{sb}). Assuming $\alpha\beta/R'\neq0$, the 1+log
condition becomes
\beq
	\alpha' = 2\frac{\beta'}{\beta} + 4 \frac{R'}{R}.
\label{dadrbR}
\eeq
This equation can be explicitly integrated,
\beq
        C e^\alpha = \beta^2 R^4,
\eeq
for some constant of integration $C$. With $\beta^2$ given in terms
of $\alpha$ and $R$ by (\ref{a2b2}), the result is
\beq
	C e^\alpha = \left(\alpha^2 - 1 + \frac{2M}{R}\right) R^4 .
\label{eqnRa0}
\eeq 
The above calculation generalizes immediately to various other lapse
conditions, including maximal slicing, harmonic slicing, or the more
general Bona-Mass\'o slicings, although in the latter case the
integration may not be possible explicitly. 
For maximal slicing, $K=0$, and the term $Ce^\alpha$ in (\ref{eqnRa0})
is replaced by $C$. For stationary
harmonic slicing, $\lie_\beta \alpha = \alpha^2 K$, and the
left-hand-side of (\ref{eqnRa0}) is $C\alpha^2$. 1+log slicing is
special in that the term $C e^\alpha$ in (\ref{eqnRa0}) introduces a
non-trivial dependence on $\alpha$. 
In \cite{HanHusOhm08,Ohm07}, these different slicings are obtained as
special cases of a more general slicing condition.

The integrated 1+log equation (\ref{eqnRa0}) does not
explicitly depend on $r$ but only on $\alpha(r)$ and $R(r)$, i.e.\ it
is the same equation whether we introduce the Schwarzschild radial
gauge $R=r$ or not, and it is equally valid for isotropic
gauge. Also, we did not have to specialize to $R=r$ to
perform the integration.

\subsection{The critical point, determining $C$}

The constant of integration $C$ is determined by requiring regularity of
$\alpha'$ for $0\leq\alpha<1$. Indeed, although integrating
(\ref{dadrbR}) when written in terms of $\beta$ is straightforward, we
have to note a regularity issue for the right-hand-side. 
Since $\alpha'(R) = \alpha'(r)/R'(r)$, substituting
for $\beta$ in (\ref{dadrbR}) gives (with $M=1$)
\beq
    \alpha'(R) = 
\frac{6 \M + 4 R (\alpha^2 - 1)}{(2 \M + R (\alpha^2 - 2 \alpha - 1))R}
\label{dadR}
\eeq
The stationary 1+log condition in this form is the starting point for
its solution in \cite{HanHusPol06,BauEtiLiu08}. 
The right-hand-side of this equation can become singular if the
denominator vanishes. We assume that $R>0$ and $R'(r)>0$, which holds
for Schwarzschild radial gauge ($R=r$, $R'(r)=1$) and which is natural
in the context of isotropic coordinates. Regularity of $\alpha'(R)$
requires that if the denominator in (\ref{dadR}) vanishes, then the
numerator has to vanish, too.
There are two solutions, one of which does fall into the interval
$0\leq\alpha<1$. The corresponding ``critical'' values
$\alpha=\alpha_c$ and $R=R_c$ are
\bea
	\alpha_c &=& \sqrt{10} - 3 \approx 0.162,
\\
	{R_c} &=& \frac{1}{4\alpha_c} =
                          \frac{1}{4}(\sqrt{10}+3) 
                          \approx 1.541.
\label{Rcritical}
\eea
The integrated 1+log equation also has to hold at the critical point
$(\alpha_c,R_c)$, which implies
\beq
        {C} = \frac{1}{2} R_c^3 e^{-\alpha_c} =
            \frac{1}{128}(3+\sqrt{10})^3 e^{3-\sqrt{10}}
            \approx 1.554.
\label{Cc}
\eeq
If we do not impose regularity at the critical point, then the
integration constant of the 1+log equation can take on different
values, which leads to various other solutions (see the discussion in
\cite{Ohm07}).

\begin{figure}[t]
\centering
\includegraphics[width=65mm]{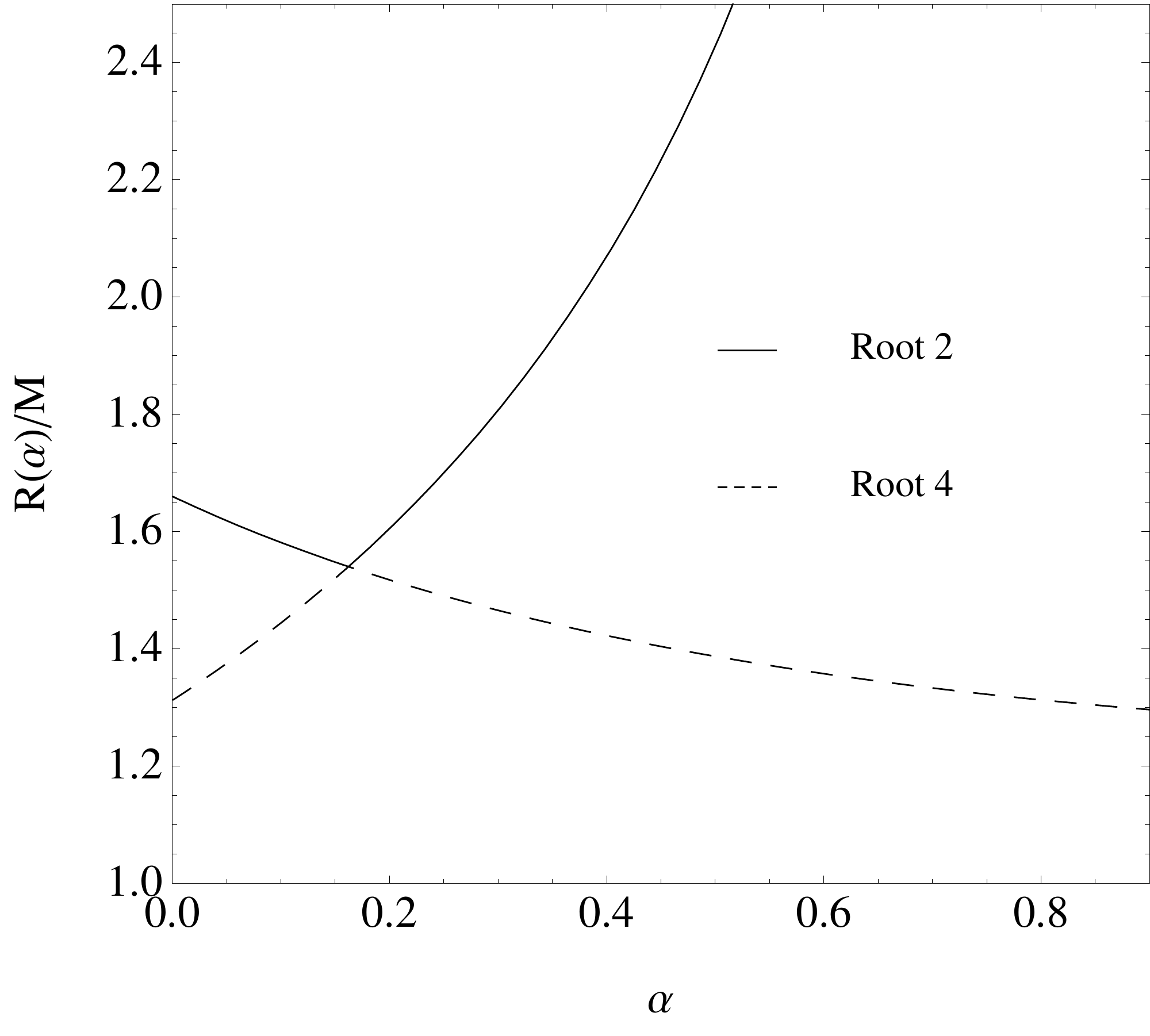}
\hspace{5mm}
\includegraphics[width=65mm]{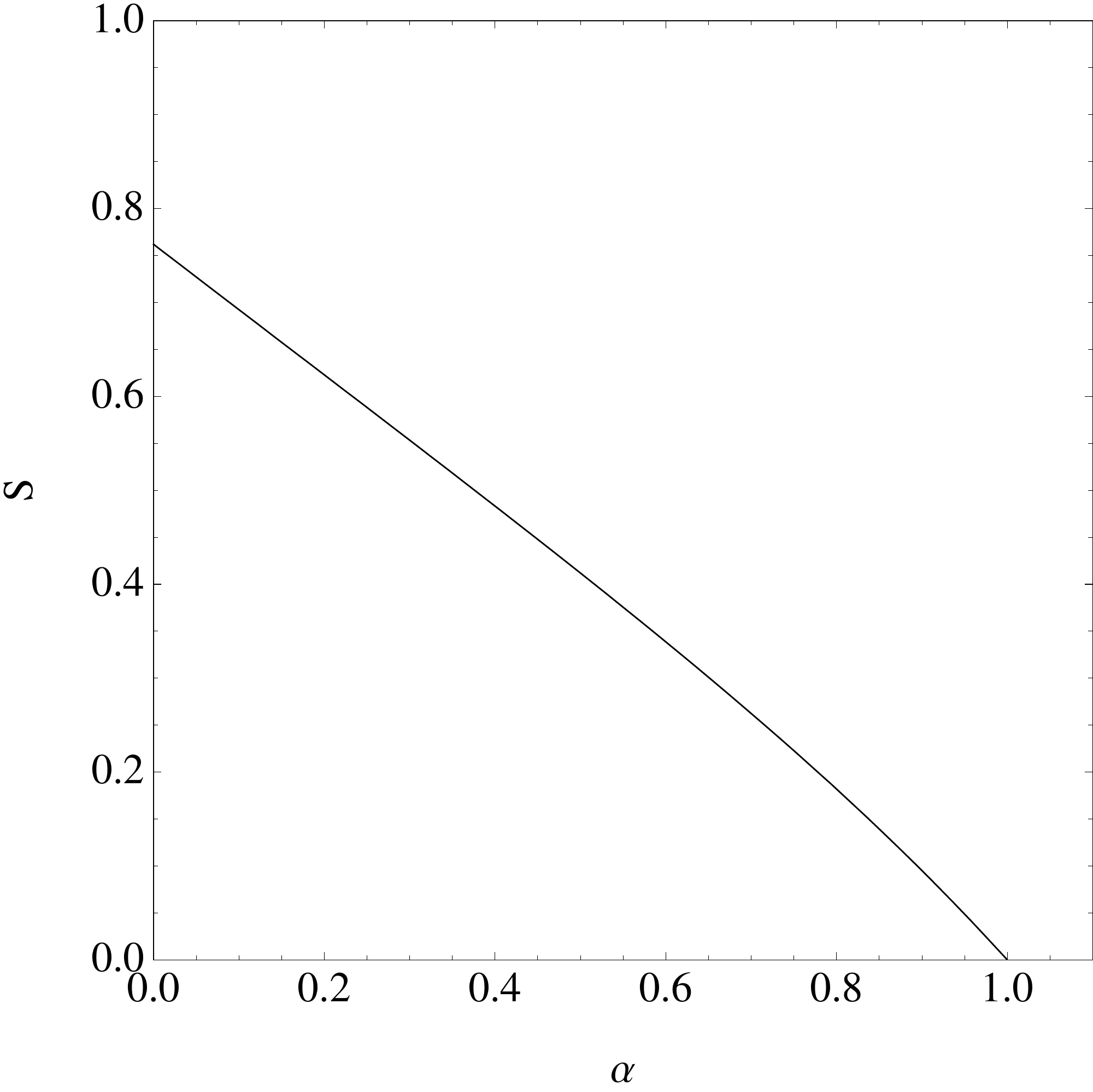}
\caption{
The two real-valued roots of the integrated 1+log equation, with
labeling introduced by Mathematica in one particular example (top panel). The
standard stationary 1+log slice corresponds to the curve that starts
at $\alpha=0$ and $R_0\approx 1.312M$, passes smoothly through
$\alpha_c\approx0.162$ and $R_c\approx1.541M$, and continues
for $\alpha\rightarrow1$ to $R\rightarrow\infty$. The other smooth branch is
ruled out by the boundary conditions.
For the standard branch, the inverse radius, $S=1/R$, shows a rather
linear dependence on $\alpha$, running from $S_0=1/R_0$ to 0 (bottom panel).
}
\label{plotRofa24}
\end{figure}

We can evaluate $\alpha'(R)$ at the critical point by l'Hopital's
rule. Taking derivatives of the numerator and denominator in
(\ref{dadR}) results in a quadratic equation for $\alpha'(R_c)$ with
the two solutions
\beq
\alpha'(R_c) = 
\frac{8 \left(-2\pm\sqrt{10+3\sqrt{10}}\right)}{16+5 \sqrt{10}}
\approx 0.607, -1.613.
\label{dadRc}
\eeq
Therefore, if we integrate (\ref{dadR}) as a first-order differential
equation for $\alpha(R)$ starting at $R=R_c$, there are two possible
initial tangents $\alpha'(R_c)$, which lead to two different
solutions. The first solution with $\alpha'(R_c)>0$ is the
standard stationary slice with $\alpha(R)$ monotonically increasing
from 0 to 1, while the second solution is decreasing with $R$. See
Fig.~\ref{plotRofa24}.

\subsection{Properties of $R(\alpha)$ and $\alpha(R)$}

We have seen that the stationary 1+log slicing condition can be
integrated explicitly to obtain an implicit equation (\ref{eqnRa0})
for $\alpha$ as a function of $R$, which however cannot be solved
explicitly. On the other hand, the same equation can be read as a
fourth-order polynomial equation for $R$ in terms of $\alpha$,
\beq
	(\alpha^2 - 1)R^4 + 2\M R^3 - C e^\alpha = 0,
\label{eqnRa}
\eeq 
which can be solved explicitly, although there are four roots to
consider.

For $C$ determined by regularity at the critical point, (\ref{Cc}),
Mathematica finds four roots $R_{r1}$, $R_{r2}$, $R_{r3}$, and
$R_{r4}$, two of which are real for $0\leq\alpha<1$. The expressions
are somewhat unwieldy so we do not show them here.  By inspection, see
Fig.~\ref{plotRofa24}, we can define
\bea
	R(\alpha) &=& R_{r4}(\alpha) \quad 
        \mbox{for $\alpha\leq\alpha_c$}, 
\nonumber \\ 
	R(\alpha) &=& R_{r2}(\alpha) \quad 
        \mbox{for $\alpha>\alpha_c$}.
\label{explicitRofa}
\eea
The labeling of the roots and of the branches is arbitrary and only
reflects Mathematica's choices in this regard. We pick the solution
that corresponds to $\alpha\rightarrow1$ for $R\rightarrow\infty$, and
that runs smoothly through the critical point. 
It is straightforward to see that $R(\alpha)$ is $C^1$ at $\alpha_c$. 
$R(\alpha)$ given in (\ref{explicitRofa}) is continuous, and so
is its first derivative based on our discussion of $\alpha'(R_c)$ in
the previous section. Taking further derivatives of
$\alpha'(R)$ given in (\ref{dadR}) should allow us to establish
smoothness at $R_c$. 

Let us now turn to $\alpha(R)$, which is implicitly defined by
(\ref{eqnRa}) or by $R = R(\alpha)$. It is instructive to consider
the two limiting cases $\alpha\rightarrow0$ and $R\rightarrow\infty$.

For $\alpha=0$, we obtain a finite value
\beq
	R_0 = R(0), \quad \alpha(R_0) = 0.
\eeq
$R_0$ can be obtained directly from (\ref{eqnRa}) by setting $\alpha =
0$,
\beq
	R_0^4 - 2\M R_0^3 + C = 0,
\label{R0C}
\eeq
which has four solutions, still too unwieldy to show in closed form. 
For the case we consider,
\beq
	R_0 \equiv R(0) \approx 1.312 \M. 
\eeq
The other real branch gives $R_0\approx1.661 \M$, see also
Fig.~\ref{plotRofa24}.

The function $\alpha(R)$ can be Taylor-expanded around $R=R_0$,
\beq
	\alpha(R) = (R-R_0)\alpha'(R_0)  
                    + \frac{1}{2} (R-R_0)^2 \alpha''(R_0) + \ldots,
\label{anearR0}
\eeq
where the leading constant vanishes by definition, $\alpha(R_0)=0$,
the linear term is directly given by the stationary 1+log condition,
(\ref{dadR}), and further coefficients can be obtained by
differentiating (\ref{dadR}). The $n$-th derivative of $\alpha(R)$ is
obtained from (\ref{dadR}) in terms $R$ and lower order derivatives of
$\alpha(R)$, which can therefore be computed iteratively for $n=1$,
$2$, $\ldots$. The first two coefficients are
\bea
	\alpha'(R_0) &=& 
     \frac{6-4 R_0}{(2-R_0) R_0},
\label{aprimeR0} 
\\
	\alpha''(R_0) &=& -\frac{4 \left(-9 R_0^3+29 R_0^2-27
   R_0+6\right)}{(2-R_0)^3 R_0^2}.
\eea
Since $R_0\neq0$ and $R_0\neq2$, all the derivatives are regular at
$R_0$. Hence, $\alpha(R)$ possesses a regular Taylor series at
$R_0$. For example, 
\beq
   \alpha'(R_0) \approx 0.832. 
\label{aprimeR0number}
\eeq

Let us also consider the limit of large $R$. The Taylor series of
$\alpha(R)$ in $1/R$ 
begins with
\beq
	\alpha(R) = 1 - \frac{1}{R} - \frac{1}{2R^2} - \frac{1}{2R^3}
                 - (5 - 4 C e) \frac{1}{8R^4} + O\left(\frac{1}{R^5}\right).
\label{aofsmallS}
\eeq
See also Fig.~\ref{plotRofa24}. Expansion (\ref{aofsmallS}) follows from
(\ref{eqnRa}) written as
$\alpha = (1 - \frac{2}{R} + \frac{C e^\alpha}{R^4})^{1/2}$ with
$e^\alpha\simeq e$ for $\alpha\rightarrow1$. This argument
assumes that $\alpha=1+O(1/R)$. Without this assumption,
Eqn.\ (\ref{aofsmallS}) can be obtained by implicit differentiation of
(\ref{eqnRa}). 
The expansion (\ref{aofsmallS}) shows that the constant of integration
$C$ of the 1+log equation is not determined by the condition that
$\alpha=1$ for $R\rightarrow\infty$ ($C$ only enters at fourth
order). This condition already holds due to asymptotic flatness of the
Schwarzschild solution (\ref{sa}) - (\ref{sb}).

\section{Isotropic Coordinates}
\label{isotropic}

Given $R(\alpha)$ and $\alpha(R)$ as the solution of the stationary
1+log condition, we turn to the isotropy condition, which relates the
coordinate radius $r$ to the areal radius $R$ and the lapse $\alpha$.
In our coordinates, spatial isotropy is implied by $f=r/R$,
(\ref{fofrR}).
For the stationary, spherically symmetric solution we have
$\alpha=fR'$, (\ref{solalpha}), so the isotropy condition becomes 
\beq
	\alpha(r) = \frac{r R'(r)}{R(r)}.
\label{aofrR}
\eeq 
This is one ODE involving two unknown functions $\alpha(r)$ and
$R(r)$, which however are related through $\alpha=\alpha(R)$, or
equivalently $R=R(\alpha)$.

\subsection{Explicit integrals for $r(R)$ and $r(\alpha)$}

Written as
\beq
	\frac{dr}{r} = \frac{dR}{\alpha(R) R},
\label{dlnr}
\eeq
the isotropy condition leads to the integral
\beq
	r(R) = C_1 \exp \int \frac{1}{\alpha(R) R} dR.
\label{rofR0}
\eeq
A change of integration variables gives
\beq
	r(\alpha) = C_2 \exp \int \frac{1}{\alpha R(\alpha)}
\frac{dR(\alpha)}{d\alpha} d\alpha.
\label{rofa0}
\eeq 
The $C_i$ are constants of integration.

The task is to find practical methods to evaluate (\ref{rofR0}) or
(\ref{rofa0}). In contrast to maximal slicing
\cite{BauNac07}, the integration cannot be performed explicitly
(i.e.\ Mathematica does not know how, and the form of $\alpha(R)$ and
$R(\alpha)$ makes the existence of an explicit solution unlikely). 

An immediate issue with (\ref{rofR0}) and (\ref{rofa0}) is
that the integral becomes divergent at its lower and upper bounds,
$\alpha=0$ and $R=R_0$, and $\alpha=1$ and $R\rightarrow\infty$. In
\cite{HanHusOhm08,Ohm07}, we described a method that allows the accurate
numerical evaluation of (\ref{rofa0}) in Mathematica. The integral is
performed as $\int_{1}^{\alpha}$, which is split into two parts
treating the cases $\alpha<0.1$ and $\alpha>0.1$ separately. For
$\alpha>0.1$, the method is based on one partial integration that
pulls out a factor of $R^{1/\alpha}$ on the right-hand-side of
(\ref{rofa0}), so that the far limit $r/R\rightarrow1$ for
$\alpha\rightarrow1$ is directly implemented. The numerical
integration relies on Mathematica's ability to handle the singularity
in the integrand for $\alpha\rightarrow0$ automatically.

Here we discuss a modified analytic formulation that regularizes the
integral at both bounds by explicitly extracting the problematic
factors. This aids numerical evaluation because no special methods for
divergent integrals have to be employed, which we explore in
Sec.~\ref{numerics}.
Furthermore, the regularized expression allows an analytic
discussion of the limit $\alpha\rightarrow0$. Our previous method is
not convenient for such an analysis because of the singular
limit and because the integration is split into two pieces.

For large $R$, the integrand in (\ref{rofR0}) asymptotes to $1/R$,
\beq
	\frac{1}{\alpha R} \simeq \frac{1}{R}  \for
R\rightarrow\infty, \alpha\rightarrow1.
\eeq
The integral $\int \frac{1}{R}dR = \ln R$ is divergent, but
in such a manner that $\exp\int\frac{1}{R}dR=R$ and hence $r\sim
R$. We therefore rewrite (\ref{dlnr}) as
\beq
	\frac{1}{r}dr = \left(\frac{1}{\alpha R} - \frac{1}{R}\right)dR 
                        + \frac{1}{R}dR,
\label{dlnr1}
\eeq
obtaining for the integral
\beq
	r = R \exp \int_\infty^R \frac{1-\alpha}{\alpha \bar R}d\bar R,
\label{rofR1}
\eeq
where we have introduced explicit integration limits, and we have
fixed the constant of integration, $C_1=1$.
For large $R$ we have the expansion (\ref{aofsmallS}) for $\alpha(R)$, 
$\alpha \simeq 1 - \frac{1}{R}$, and the integrand now has the
asymptotic behavior 
$\frac{1-\alpha}{\alpha R}\simeq\frac{1}{R^2}$. 
The integral is convergent for a fixed $R>R_0$, and
\beq
	r \simeq R   \for R\rightarrow\infty.
\eeq

For an alternative derivation of (\ref{rofR1}), note that
a natural variable to consider is $r/R=f=\psi^{-2}$. Differentiating
$r/R$ with respect to $R$ and using the isotropy condition, we get
\beq
	\frac{d(r/R)}{r/R} = \frac{1-\alpha}{\alpha R} dR, 
\eeq
in analogy to (\ref{dlnr}), which integrates directly to
(\ref{rofR1}). In other words, scaling $r$ by $R$ avoids the detour
of regularizing a singular integral whose exponential gives the
required factor $R$ for $r\simeq R$. 

For $R$ approaching its lower limit, the integrand of the upper-limit
regularized expression (\ref{rofR1}) and the original integrand both
have the pole 
\beq
	\frac{1-\alpha}{\alpha R} \simeq 
        \frac{1}{\alpha R} \simeq \frac{1}{a_1(R-R_0)R_0}  
        \for R\rightarrow R_0, \alpha\rightarrow0,
\eeq
where we have used the leading order linear term of the expansion of
$\alpha$ at $R_0$, (\ref{anearR0}), with $a_1=\alpha'(R_0)$
given in (\ref{aprimeR0}) and (\ref{aprimeR0number}). If we subtract
this pole as we did with $1/R$ for $R\rightarrow\infty$, the lower
limit becomes regular, but the upper limit picks up the 
singularity $\ln(R-R_0)$. Instead we write (with one factor $R_0$
replaced by $R$)
\bea
   \frac{dr}{r} &=& \left(\frac{1}{\alpha R} - \frac{1}{R} -
                        \frac{1}{a_1(R-R_0)R}\right)dR 
\nonumber \\
                && + \frac{dR}{R} + \frac{dR}{a_1(R-R_0)R}.
   \label{dlnr2}	
\eea
With $\int\frac{1}{(R-R_0)R}dR = \frac{1}{R_0}\ln(1-\frac{R_0}{R})$,
\bea
    r(R) &=& R\, \left(1 - \frac{R_0}{R}\right)^\gamma \exp I(R),
\label{rofR2}
\\
   I(R) &=& 
        \int_\infty^R 
        \left(\frac{1-\alpha}{\alpha} - \frac{\gamma R_0}{\bar R-R_0} \right) 
        \frac{d\bar R}{\bar R},
\label{IofR}
\eea
where $\gamma = 1/(a_1 R_0)$, or with (\ref{aprimeR0}),
\beq
	\gamma = \frac{2-R_0}{6-4R_0} \approx 0.916.
\label{gamma}
\eeq
Our final expression for $r(R)$, (\ref{rofR2}), involves an integral 
$I(R) = \int_\infty^R(\ldots)$. The goal was to remove all singular
terms from the integrand and the integral, and it is straightforward
to see that $I(R)\simeq0$ for $R\rightarrow\infty$ and that $I(R_0)$
is finite, although $I(R_0)$ is not explicitly available. The
numerical result is that $\exp I(R)$ varies monotonically between
about $1.155$ and $1$, see Sec.~\ref{isodiscuss}. 

Given $r(R)$, we can compute $r(\alpha)$ as $r(R(\alpha))$. Another
possibility is to start with (\ref{rofa0}) and remove the singularity
from the integrand as we did for (\ref{rofR0}), which results in
\bea
	r(\alpha) &=& R(\alpha) \alpha^\gamma \exp J(\alpha),
\label{rofa2}
\\
	J(\alpha) &=& \int_1^\alpha
\left((1-\bar\alpha)\frac{R'(\bar\alpha)}{R(\bar\alpha)} - \gamma\right)
	\frac{d\bar\alpha}{\bar\alpha}.
\eea 
Here
$R'(\alpha)=1/\alpha'(R)$, so that from the
1+log condition (\ref{dadR})
\beq
	\frac{R'}{R}(\alpha) = 
	\frac{2+R(\alpha^2-2\alpha-1)}{6+4R(\alpha^2-1)},
\quad
	\frac{R'}{R}(0) = \frac{2-R_0}{6-4R_0} = \gamma.
\label{RpRofa}
\eeq
Again, we find that the integrand of $J(\alpha)$ is regular,
$J(\alpha)$ is finite, and $\exp J(\alpha)$ is of order 1.
When comparing $r(R)$, (\ref{rofR2}), and $r(\alpha)$, (\ref{rofa2}), for
$R\rightarrow R_0$ and $\alpha\rightarrow0$, there is an additional
factor of $R\simeq R_0$.

\subsection{ODE integration for $\alpha(r)$}
\label{isoodeint}

The preceding section shows how to obtain explicit integrals for
$r(\alpha)$, although e.g.\ for initial data we want to compute the
inverse relation $\alpha(r)$. Using the chain rule,
\beq
	\frac{d\alpha(r)}{dr} = \frac{d\alpha(R)}{dR} \frac{dR(r)}{dr}.
\eeq
The first factor on the right-hand-side is given by the
(non-integrated) 1+log condition (\ref{dadR}), the second factor by the
isotropy condition (\ref{aofrR}). This equation determines $\alpha(r)$ by
an ODE,
\beq
	\frac{d\alpha}{dr} = \frac{\alpha}{r} \frac{R(\alpha)}{R'(\alpha)},
\label{dadr}
\eeq
where $R'/R$ is given in terms of $R(\alpha(r))$ and $\alpha(r)$ in
(\ref{RpRofa}).  Previously, we integrated this equation as $\int
dr/r$ obtaining $r(\alpha)$, cmp.\ (\ref{rofa0}). 
Incidentally, writing this as $\int d\alpha/\alpha$ does not give us
$\alpha(r)$ directly, since we do not have $R(r)$ available and
therefore cannot perform the integration directly. 

For $R\rightarrow R_0$ and $\alpha\rightarrow0$, the isotropy
condition in the form $\alpha = r R'(r)/R$ implies that
$r\rightarrow0$ or $R'(r)\rightarrow0$. We consider
$r\rightarrow0$. 
In this limit, we can evaluate $\alpha'(r)$ by using l'Hopital's rule
on the right-hand-side of (\ref{dadr}). Since
$\frac{\alpha}{r}\rightarrow\alpha'(0)$, we find that
\beq
	\alpha'(0) = \alpha'(0) \frac{6-4R_0}{2-R_0}, 
\label{aofr0}
\eeq
compare (\ref{RpRofa}). A priori there are two possibilities.
If $\alpha'(0) = 0$, then (\ref{aofr0}) is satisfied for all
$R_0\neq2$. If $\alpha'(0) \neq 0$, then we conclude that $R_0 =
\frac{4}{3} \approx 1.333$, which accidentally is rather close to 
$R_0 \approx 1.312$ determined by the regularity condition at $R_c$. 

Therefore, if we start integrating the combined isotropy and 1+log
condition (\ref{dadr}) at $r=0$, there is one distinguished case that
is singled out by the assumption that $\alpha(0)=0$ and $\alpha'(0)\neq0$,
e.g.\ $\alpha(r)= a_1 r + a_2 r^2 + \ldots$ with $a_1\neq0$. This
implies $R_0 = \frac{4}{3}$. Only when integrating the ODE for
$r\rightarrow\infty$ do we discover that these slices do not pass
through $(\alpha_c,R_c)$ and do not have the appropriate far limit. 

Eqn.\ (\ref{aofr0}) for $\alpha'(0)=0$ is, however, consistent with the
ansatz
\beq
	\alpha(r) \sim r^{1/\gamma},
\label{aofrsim}
\eeq
consistent with the calculation of $r(\alpha)$ leading to
(\ref{rofa2}). In other words, the leading order behavior of
$\alpha(r)$ near $r=0$ can be directly obtained from the ODE
(\ref{dadr}), if we use the information about global properties
obtained for $\alpha(R)$, in particular that $R_0\neq\frac{4}{3}$.    

The constant of proportionality in (\ref{aofrsim}) is not determined
locally but requires knowledge of a global integral and depends on the
critical constant $C$. Although we cannot invert (\ref{rofR2}) and
(\ref{rofa2}) to obtain $R(r)$ and $\alpha(r)$ explicitly for all $r$,
from (\ref{rofR2}) and (\ref{anearR0}) for $r\rightarrow 0$ we obtain
\beq
	\alpha(r) \simeq \frac{1}{\gamma} e^{-I(R_0)/\gamma} 
        \left(\frac{r}{R_0}\right)^{1/\gamma}.
\eeq

\subsection{Some quantitative results for isotropic coordinates}
\label{isodiscuss}

\begin{figure}[t]
\centering
\includegraphics[width=65mm]{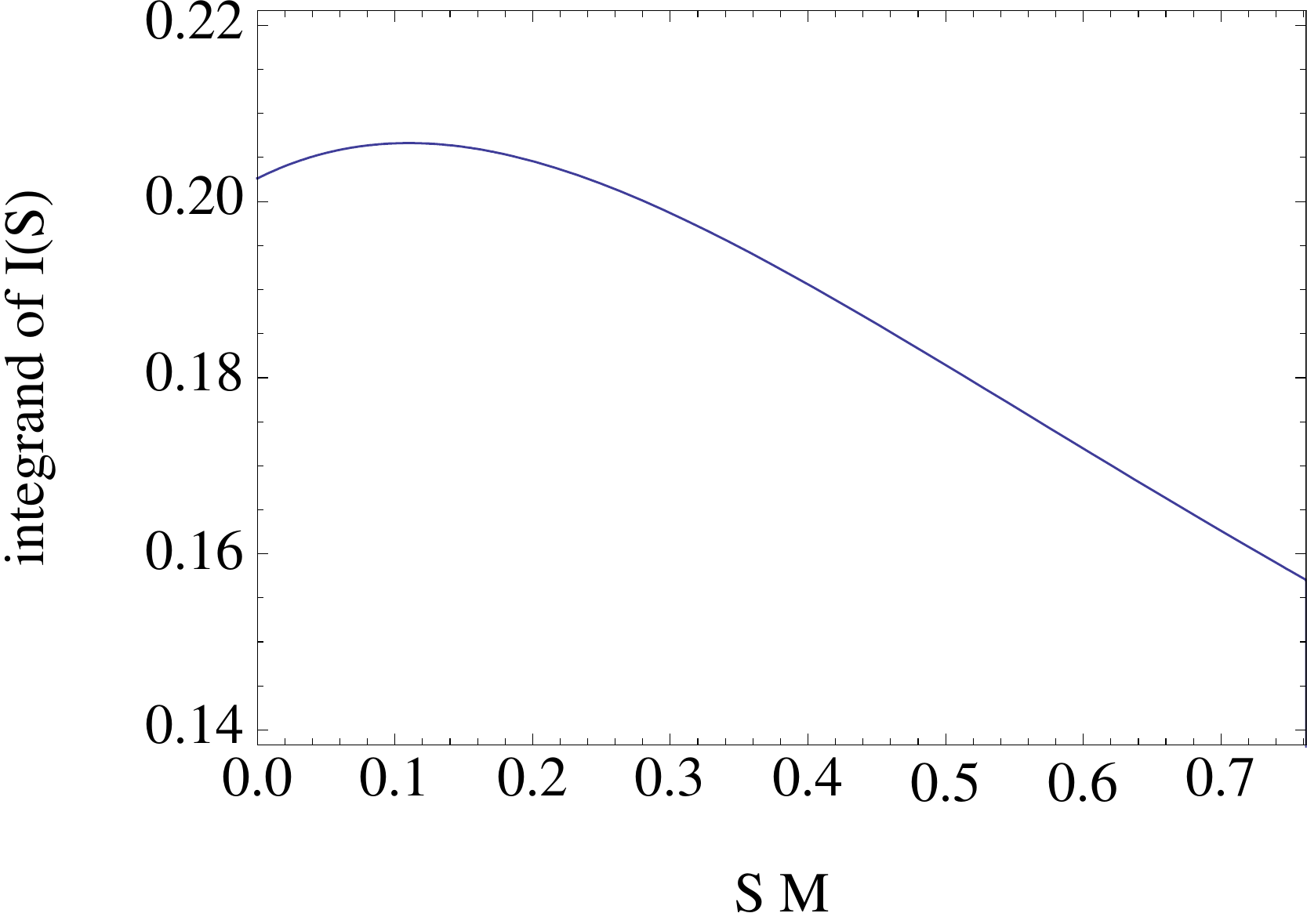}
\quad\quad 
\includegraphics[width=65mm]{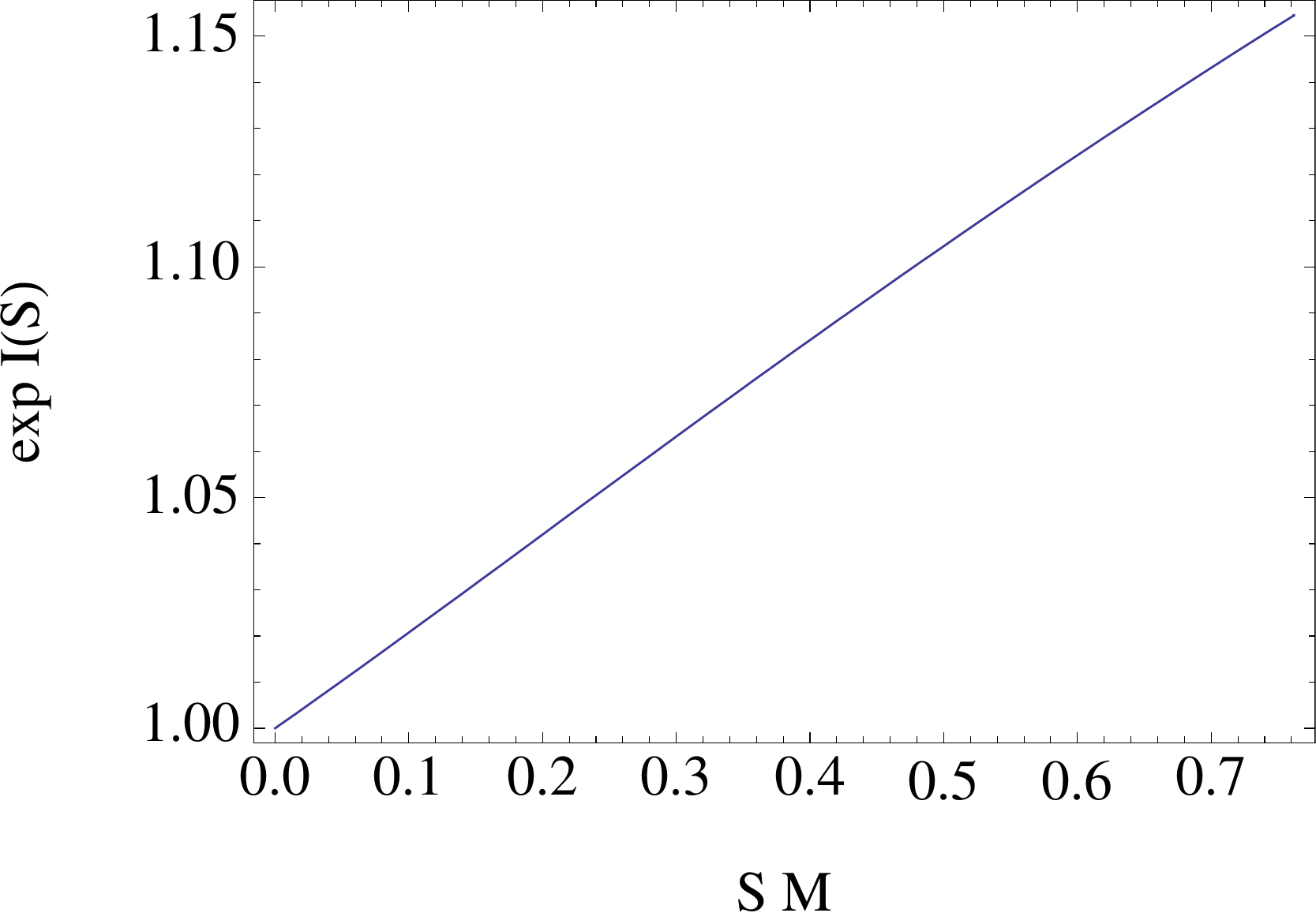}
\caption{
The regularized integrand and its exponentiated integral $\exp I(S)$
ocurring in the integration formula for the isotropic radius $r$,
plotted versus $S=1/R$ from $S=0$ to $S_0=1/R_0\approx0.762/M$. The
resulting effect in $r(R)$ is a factor
of order unity that varies monotonically from 1 at $R=\infty$ to
$1.155$ at $R=R_0$.  
}
\label{plotIofS}
\end{figure}

Fig.~\ref{plotIofS} shows numerical results for the integral
involved in the calculation of $r(R)$ based on (\ref{rofR2}) and
(\ref{IofRusingS}).
The main feature is that $\exp I(R)$ varies monotonically and almost
linearly between
\beq
	\exp I(R_0) \approx 1.155, \quad \exp I(\infty) = 1.
\eeq
Considered for $R\rightarrow R_0$,
\beq
	\exp I(R) = \exp I(R_0) \left[1 + (R-R_0) I'(R_0) + \ldots\right],
\eeq
where $I'(R_0) = - \frac{1}{R_0} (1 + \frac{a_2}{a_1^2})$ with
$a_1=\alpha'(R_0)$ and $a_2=\alpha''(R_0)$.

Since $\exp I(R)$ is of order unity,
$r(R)$ is approximated by 
\beq
	r \approx R (1 - \frac{R_0}{R})^\gamma = 
                  R^{1-\gamma} (R - R_0)^\gamma
\label{rofRapprox}
\eeq 
For comparison, the isotropic radius $r_w$ of wormhole, fixed-puncture data
is given by $R = (1+\frac{1}{2r_w})^2 r_w$, or
\beq
	r_w = \frac{1}{2} 
              \left(R - 1 \pm R^\frac{1}{2} (R-2)^\frac{1}{2}\right). 
\eeq
Apart from the symmetry corresponding to the wormhole, there is some
structural similarity to (\ref{rofRapprox}), i.e.\ one could define
$\gamma_w=\frac{1}{2}$ and $R_{0w}=2$. 

For $R\rightarrow\infty$, we have $r\simeq R$, while for $R\rightarrow R_0$,
\bea
	r(R) &\simeq& R^{0.084} (R-R_0)^{0.916} \exp I(R_0) 
\nonumber \\
             &\simeq& 1.181 (R-R_0)^{0.916}, 
\eea
with $r(R_0) = 0$.
Therefore, since $\gamma\approx0.916$ happens to be rather close to
unity, and since $\exp I(R_0)$ is only slowly varying, we expect
$r(R)$ to be well approximated by the linear term $(R-R_0)^1$ over the
entire range of $R$.

\begin{figure}[t]
\centering
\includegraphics[width=70mm]{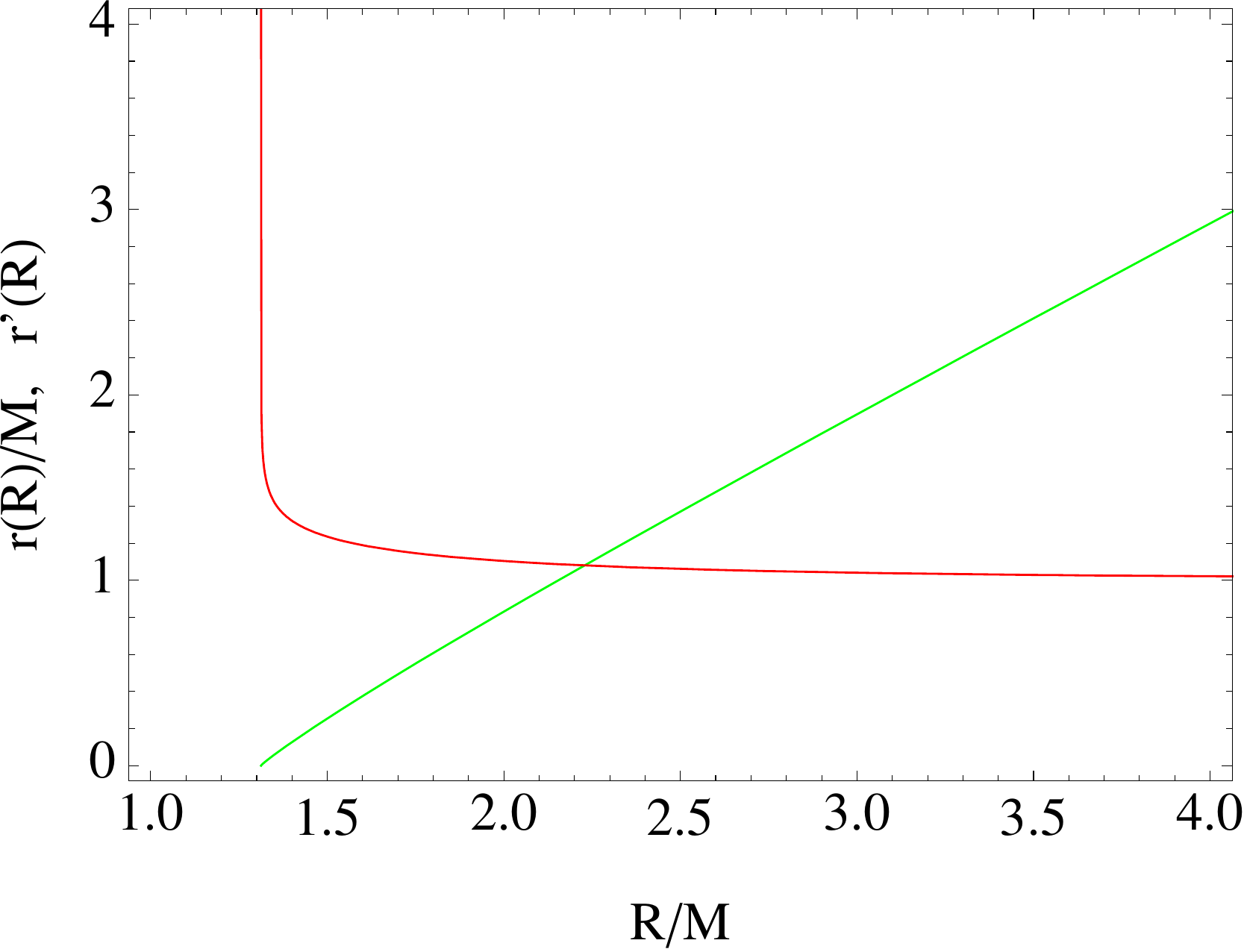}
\caption{
The isotropic radius $r(R)$ and its first derivative $r'(R)$. As
expected, $r(R)$ is approximately linear for $R$ varying from $R_0$ to
$\infty$. Its first derivative has a weak pole singularity at $R_0$
with exponent $\gamma-1\approx-0.084$. 
}
\label{plotranddrdR}
\end{figure}

\begin{figure}[t]
\centering
\includegraphics[width=65mm]{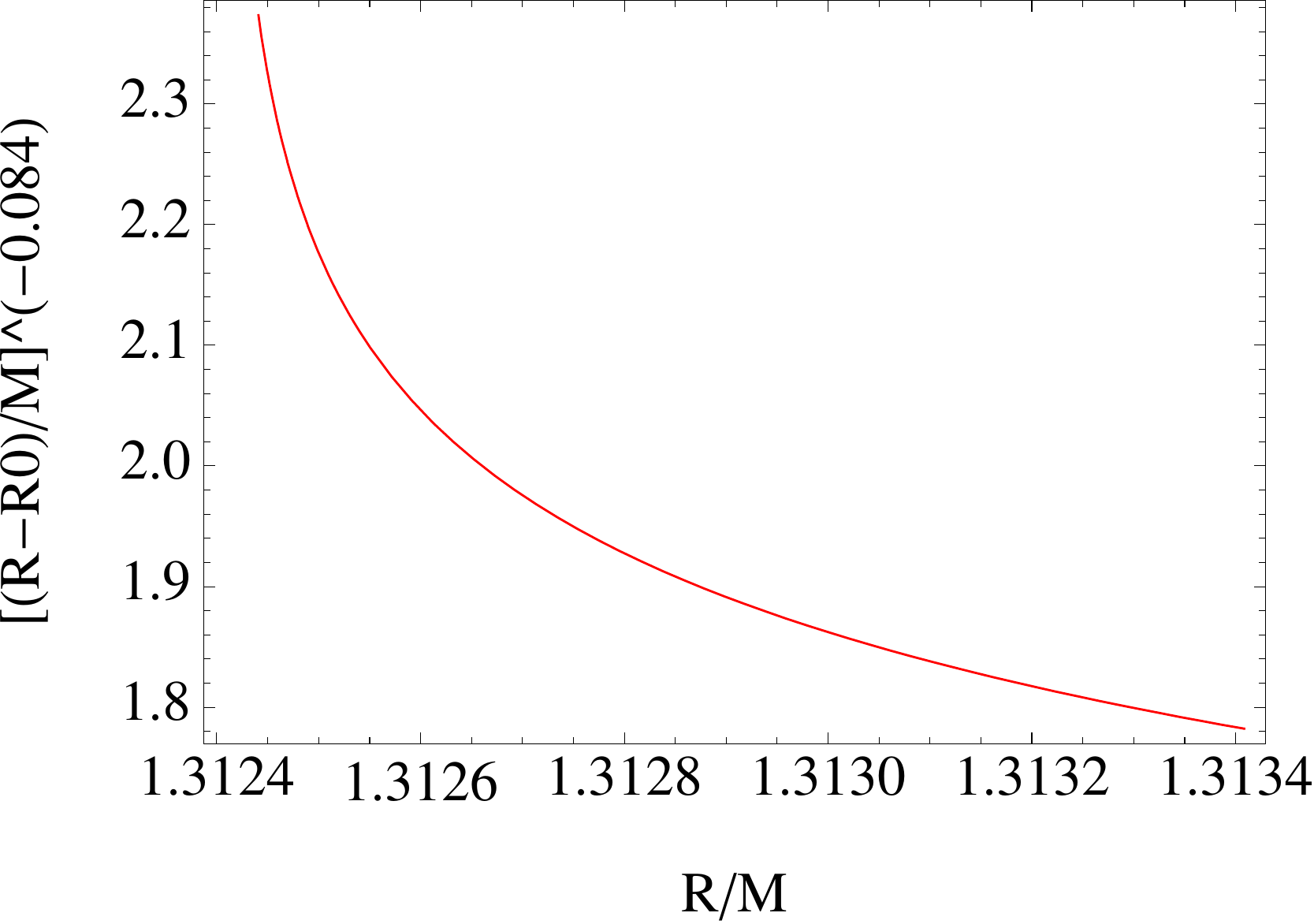}
\quad\quad
\includegraphics[width=65mm]{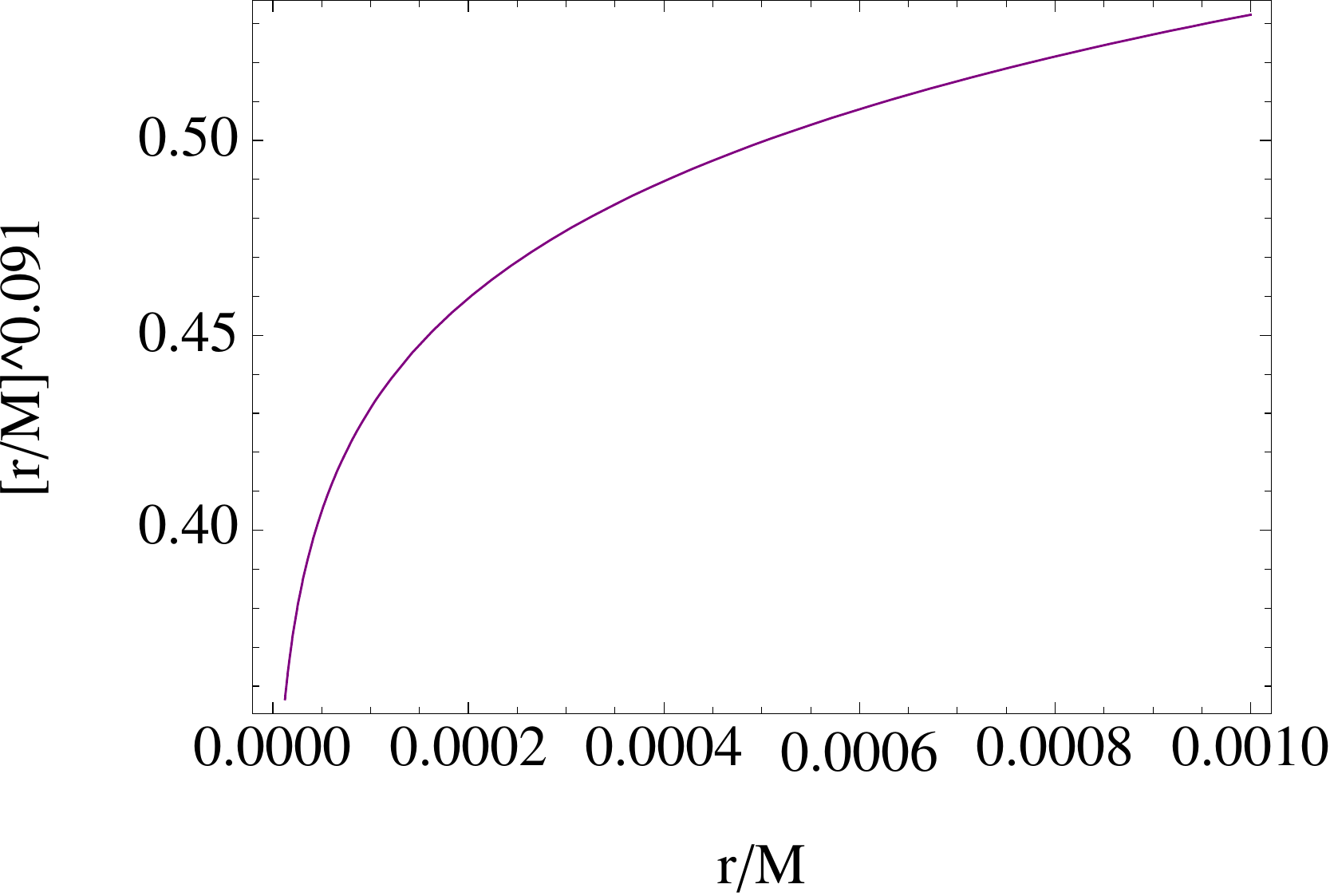}
\caption{
Singular behavior of $(R-R_0)^{\gamma-1}\approx (R-R_0)^{-0.084}$ for R in the range from $R_0$ to $R_0+10^{-3}M$ (top panel), and vanishing of $r^{-1+1/\gamma}\approx r^{0.091}$ near $r=0$ (bottom panel). 
}
\label{plotdrdRzoom}
\end{figure}

Fig.~\ref{plotranddrdR} shows the quantitative result for $r(R)$,
and also for $r'(R)$. Since $\gamma<1$, the first derivative has a
weak pole singularity at $R_0$, 
\beq
	r'(R) \sim (R-R_0)^{\gamma-1} \approx 
	\left( \frac{1}{R-R_0} \right)^{0.084}.
\label{rofRsing}
\eeq
The same singularity occurs in $r'(\alpha)$, the first derivative of
$r(\alpha)\sim \alpha^\gamma$, see (\ref{rofa2}). As a quantitative
measure of the singularity, we plot (\ref{rofRsing}) in
Fig.~\ref{plotdrdRzoom}. As another example, $1/x^{0.084}=10$ implies
$x\approx1.1\times10^{-12}$.

A numerical code using isotropic coordinates typically does not compute
$r(\alpha)$, but for example
\beq
	\alpha(r) \sim r^{1.091}, \quad
        \alpha'(r) \sim r^{0.091}, \quad
        \alpha''(r) \sim r^{-0.909}.
\label{dadrexamples}
\eeq
If a numerical code relied on $\alpha'(0)=0$ at $r=0$, then some inordinate
amount of resolution would be required, see Fig.~\ref{plotdrdRzoom}.
However, typically even $\alpha''(r)$ is regularized properly in a
standard BSSN moving puncture code due to the factor of $\psi^{-4}\sim r^2$
multiplying the second derivative of the lapse in the evolution
equations for the extrinsic curvature. The examples
(\ref{dadrexamples}) show that the lapse is slightly ``more regular''
at $r=0$ than $\alpha(r)\sim r$.


\section{Numerics}
\label{numerics}

For various applications of the stationary 1+log slices there remains
the task to find numerical solutions for $R(\alpha)$ and $\alpha(R)$,
and to compute the integrals of the coordinate transformation to
isotropic coordinates to obtain $R(r)$ and $\alpha(r)$. We comment on
suitable numerical methods in this section.

\subsection{Numerical computation of $R(\alpha)$ and $\alpha(R)$}
\label{numcompRa}

An explicit form for $R(\alpha)$ is readily obtained based on the
roots of a fourth order polynomial, see
(\ref{explicitRofa}). Implemented in Mathematica, we may have to
increase the default working precision for $\alpha$ close to 1, and we
have to remove a small complex part of the result at round-off level.

When working outside a package like Mathematica (for concreteness
think of an implementation in C or C++), a simpler strategy is to work
with a general root finding routine, for example the bracketed Newton
method of \cite{PreFlaTeu92}.  This allows us to find both $R(\alpha)$ and
$\alpha(R)$, as well as inverting other implicit relations, i.e.\ we
typically need such an algorithm anyway. We save the work of
implementing the explicit but lengthy formula produced by
Mathematica. Newton's method typically requires only 3 to 6
iterations to reach double precision, taking less time than evaluating
Mathematica's formula (although there are more compact formulas
available).

\begin{figure}[t]
\centering
\includegraphics[width=65mm]{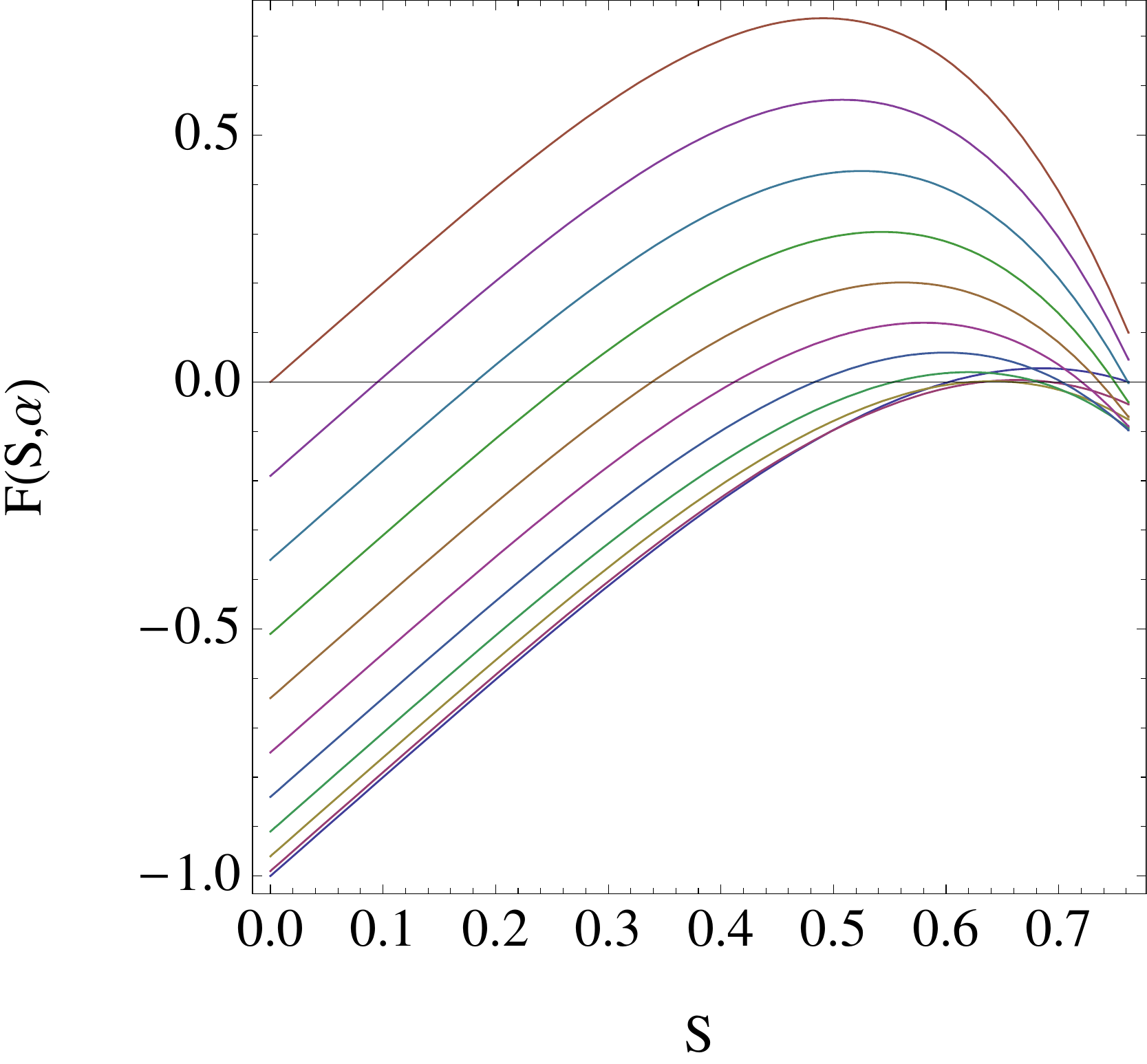}
\quad\quad 
\includegraphics[width=65mm]{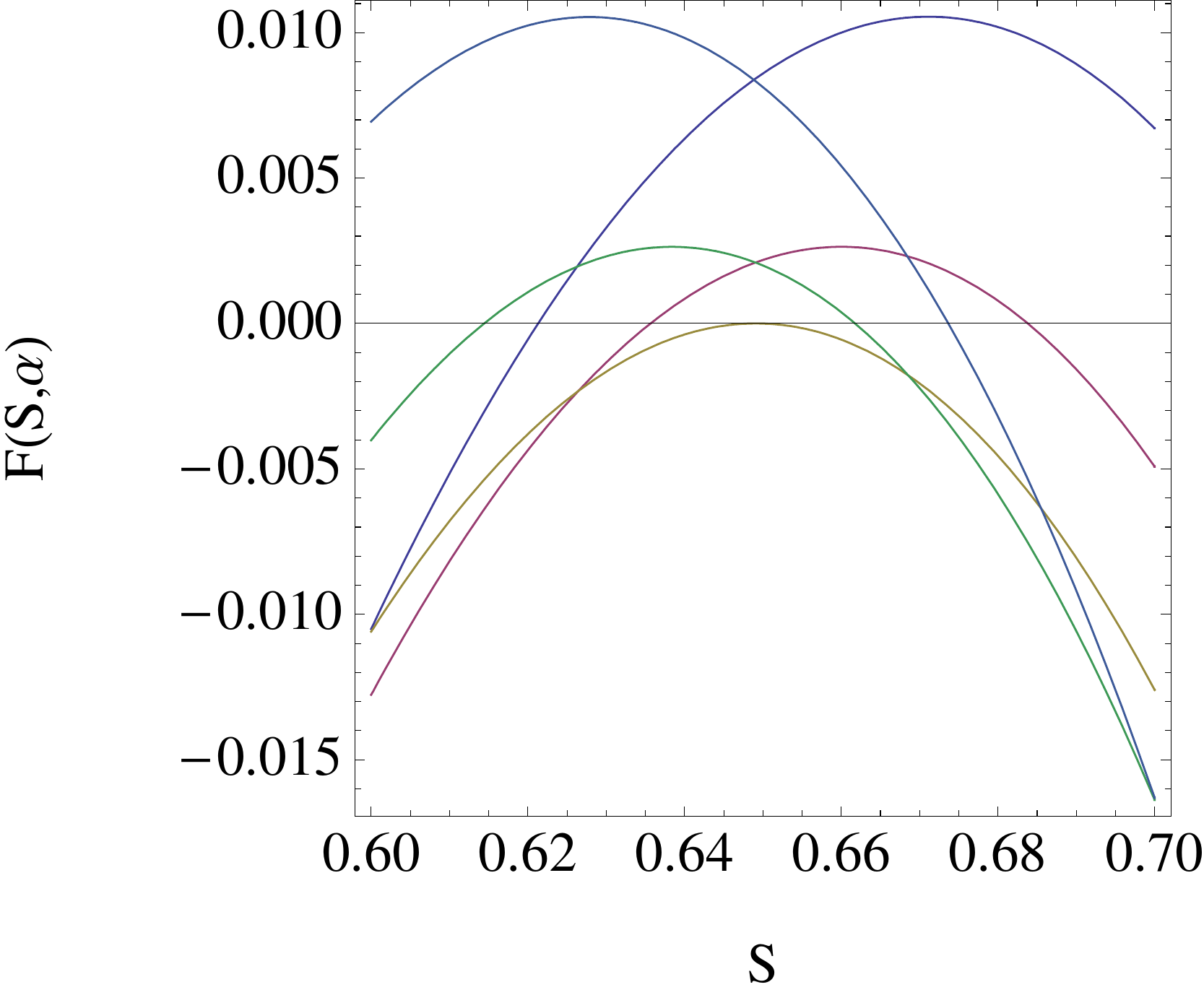}
\caption{
Solving the integrated stationary 1+log condition by root finding. Shown is
$F(S,\alpha)$ as a function of $S=1/R$ for $\alpha=0.0$ to $1.0$ in
steps of $0.1$ (top panel) and a zoom-in for $\alpha=\alpha_c-0.10$
to $\alpha_c+0.10$ in steps of $0.05$ (bottom panel). The maximal value
of $S$ is $S_0\approx0.762$, the critical value is $S_c\approx0.649$.
The solution $S(\alpha)$ is defined by $F(S,\alpha)=0$, and there may
be one or two solutions for a given $\alpha$. The root can be
bracketed by considering $[0,S_c]$ if $\alpha\in[\alpha_c,1]$ and
$[S_c,S_0]$ if $\alpha\in[0,\alpha_c]$.
}
\label{plotFofS}
\end{figure}

To implement Newton's method, it is convenient to introduce $S=1/R$,
which transforms $R\in[R_0,\infty)$ to the finite interval $S\in[S_0,0)$.
The integrated 1+log condition gives
\bea
   F(S,\alpha) &=& \alpha^2 - 1 + 2S - C e^\alpha S^4
\label{newtonF}
\\
   \frac{\partial F}{\partial S} &=& 2 - 4 C e^\alpha S^3
\label{newtondFdS}
\\
   \frac{\partial F}{\partial \alpha} &=& 2\alpha - C e^\alpha S^4
\label{newtondFda}
\eea
For example, to obtain $S(\alpha)$, fix $\alpha$ and iterate $S$ by
Newton's method for $F(S,\alpha)=0$ with derivative
(\ref{newtondFdS}).
The bracketed Newton method of \cite{PreFlaTeu92} combines the standard Newton
method with a bisection method that is used whenever the Newton step
appears to fail, and which in particular avoids that the iteration
leaves a given interval. To initialize the method, we have to localize
the root approximately, see Fig.~\ref{plotFofS}. Although
$S(\alpha)$ and $R(\alpha)$ are monotonic, $F(S,\alpha)$ as a
function of $S$ is not. We can bracket the root by setting the
starting interval to $[0,S_c]$ if $\alpha\in[\alpha_c,1]$ and to
$[S_c,S_0]$ if $\alpha\in[0,\alpha_c]$, where
$S_c=1/R_c$. Mathematica's FindRoot routine applied to the problem
requires similar care (i.e. it fails if the root is not bracketed correctly). 
We also found that Newton's method works rather well for
$R$ on the unbounded interval, for which we can obtain upper and
lower bounds for the root from (\ref{aofsmallS}). Simple bisection
does the job, too, giving on average the expected three digits for ten
iterations.

In practice, root finding works well and is applicable for both
$R(\alpha)$ and $\alpha(R)$. It can be straightforwardly and
efficiently implemented using a simple Newton method
\cite{PreFlaTeu92} (and also with Mathematica's FindRoot), the only
issue being that we have to initialize differently for
$\alpha<\alpha_c$ and $\alpha>\alpha_c$.

\subsection{Numerical integration for isotropic coordinates}
\label{isonumint}

Studying the near and far limit of the isotropy condition has left the
integrals $I(R)$ or $J(\alpha)$ for numerical evaluation. Although the
integrands are regular, at the boundaries they are defined by
one-sided limits, which is sometimes called ``regular improper''
boundaries. In practice it is often sufficient to apply ``open''
integration formulas, for which the integrands are required close to
but not at the boundaries. Accuracy can deteriorate close to an
improper boundary, and if this is an issue, Taylor expansions could be
used.  As already pointed out, if only a numerical answer is required,
we can rely on numerical integration routines as available in
Mathematica that actually handle improper as well as some types of
singular boundaries. In this case the transformation to regular
boundaries is not always necessary, cmp.~\cite{HanHusOhm08}.

Such routines also handle integrating to $\infty$, as
required in $I(R)$. As an alternative, we can change the integration
variable to $S=1/R$,
\beq
	I(R) = - \int_0^{1/R}
       \left(\frac{1-\alpha}{\alpha} - \frac{\gamma S}{S_0-S} \right) 
        \frac{d S}{S}.
\label{IofRusingS}
\eeq
This formula can be directly evaluated using e.g. Romberg integration
\cite{PreFlaTeu92}.

The domain of integration in $J(\alpha)$ is finite, but in this case
there is a potential issue at $\alpha_c$, where the regularity of
$R'(\alpha)$ relies on the cancellation of roots in
the numerator and denominator. For the purpose of integration, split
the integral into two pieces as required to handle $\alpha_c$ as an
improper boundary. (If this is not done explicitly, the Romberg integration
routine may place integration points randomly close to $\alpha_c$,
leading to unexpectedly large errors there.)

When computing initial data in isotropic coordinates, $r(R)$ and
$r(\alpha)$ actually still have to be inverted to obtain $R(r)$ and
$\alpha(r)$. This can be achieved by root finding similar to what was
discussed in Sec.~\ref{numcompRa}, so we do not discuss this here in
further detail.

Integrating (\ref{dadr}) as an ODE gives an alternative
method to obtain $\alpha(r)$ directly. Examples for useful general
purpose integrators are Mathematica's NDSolve or the Bulirsch-Stoer
routine described in \cite{PreFlaTeu92}. From the point of view of using
a minimal number of numerical algorithms, the ODE integration can be
traded for Romberg integration or whatever else is used to obtain
$J(\alpha)$. Also, in this case we can avoid using a root finding
routine if we implement $R(\alpha)$ by explicit root formulas.
Integrate (\ref{dadr}) for $\alpha(r)$ by some ODE algorithm using
the explicit root for $R(\alpha)$ on the right-hand-side. 
Given $\alpha(r)$, compute $R(r)=R(\alpha(r))$ using explicit roots.
All other quantities are then directly available in terms of $\alpha(r)$ and
$R(r)$. 
  
The issue with integrating an equation for $\alpha(r)$ is that we need
a convenient starting point. Both $r=0$ and $1/r=0$ are improper but
regular limits. A standard way to proceed is to factor the singularity
and use Taylor expansions to integrate away from the boundary. For
$\alpha(r)$ (as opposed to $R(r)$), we also have to handle the
critical point at $r_c$. Note that $r_c$ is not directly available, so
we cannot start the integration there without, say, performing the
explicit integrals discussed above. On the other hand, since accuracy
near $r_c$ is crucial for a continuous derivative of $\alpha(r)$,
starting at $r_c$ seems the most promising strategy.

To end with a concrete suggestion, integrate 
\beq
	S'(r) = - \frac{S(r)\alpha(S(r))}{r}
\eeq
as an ODE for $S(r)$, letting the automatic step size control of the
integrator handle the limits. Compute $\alpha(S)$ by Newton's method
from (\ref{newtonF}). To avoid issues with starting at improper
boundaries, start the integration at
\beq
	r_c = 0.30345204271479997 \M, \quad S(r_c) = 4(\sqrt{10}-3),
\eeq
cmp.~(\ref{Rcritical}).
That is, there is one magic number which we provide for the
convenience of the reader based on the explicit integrals,
$r_c=r(R_c)$ using Mathematica with 20 digit accuracy. Given $S(r)$,
compute $R(r)=1/S(r)$, $\alpha(r)=\alpha(S(r))$ and
$f(r)=\psi^{-2}(r)=r/R(r)$.

Independently of how the relation between $r$ and $R$ was obtained,
all other quantities required to specify conformally flat initial data
in ADM form follow from (\ref{solbeta})--(\ref{soll}).


\section{Discussion}
\label{discussion}

We have computed the stationary 1+log slices of the Schwarzschild
solution in isotropic coordinates. The computation goes beyond what
was already known by providing alternative integration methods of the
isotropy condition that simplify numerical integration, and by giving
direct access to local expansions in the isotropic radius $r$ near
$r=0$.

Let us emphasize that the 3D numerical evolutions do {\em not} use
isotropic coordinates. Specifically, even though $\psi$ is chosen to
obtain a metric with uniform determinant, the metric is not diagonal
and the conformal metric components can easily deviate by $15\%$ or
$20\%$ from $1$. The reason is that although the initial data is
conformally flat, there is some significant initial gauge evolution in
which e.g.\ the shift evolves from identically zero to the stationary
Gamma-freezing shift. During this time the coordinates are in
motion, and the final transformation between the coordinate $r$ and
the areal $R$ depends on details of the shift condition and e.g.\ also
the initial value of the lapse. For example, the size of the shift
damping parameter $\eta$ directly influences the final gradients in
$R(r)$. See \cite{Bro07a,HanHusOhm08} for the rather involved
procedure to transform the non-isotropic, numerical coordinates to
some standard coordinates.  

For the analytic stationary 1+log solution in isotropic coordinates,
we showed that near the puncture factors of $r$ as well as
$r^{1/\gamma}$ play a role. Concretely, $\psi^{-2}\sim r$ and
$\beta^r\sim r$, while $\alpha\sim r^{1.091}$. 

This was not detected in the numerical 3D data since we did not search
for a small deviation from $\alpha\sim r^{1.000}$ like
$\alpha\sim r^{1.091}$. The main open question in this regard is
whether and how much the numerical gauge (which as pointed out is not
isotropic) affects these exponents.  What are the deviations from
isotropy, do they matter at the puncture? In the original ansatz,
$g_{ij} = \delta_{ij} + O(r^2)$, implying that locally at the puncture
the coordinates are isotropic. If this holds for arbitrary gauge to
leading order, the analysis of isotropic coordinates leading to
$\alpha\sim r^{1/\gamma}$ applies also to the numerical simulations.

Furthermore, if stationary 1+log slices in isotropic coordinates are
chosen as initial data, then our analysis of the coordinate
singularity at the puncture can answer the question what singularity
the numerical evolution methods have to handle, if the non-advected
Gamma-freezing shift condition is used that preserves the
isotropy. The advected Gamma-freezing shift condition leads to
different radial coordinates, which in principle could also be
analyzed along the lines discussed here for the isotropic case.
(See \cite{GunGar06,MetBakKop06} for different types of advection 
in the shift condition using either
$\partial_t$ or $\partial_t-\beta^i\partial_i$ as time derivative.)

Based on our discussion on
integrating the 1+log equation for $\alpha'(r)$ in
Sec.~\ref{isoodeint}, one point to make is that a local Taylor
expansion at $r=0$ cannot take into account global boundary conditions
or the regularity condition at $\alpha_c$. Given any starting value for
$R_0$ for $r(R_0)=0$, we can integrate $r'(R)$ towards larger $R$, but
only specific choices lead to the standard slice (compare \cite{Ohm07}). For
the standard slice, we determine $R_0$ from (\ref{R0C}) as a function
of $C$, which we set to its critical value in (\ref{Cc}) based on
regularity at $R_c$. If we choose a different $R_0$, we get a
different slice. Expanding the BSSN equations and the gauge conditions
near $r=0$ assuming $\alpha\sim r$ rather than $\alpha\sim
r^{1/\gamma}$ therefore may well lead to a consistent local solution,
which however is incompatible with the global solution we are looking
for. 
More generally, it would be relevant to determine which leading terms
of the expansion are independent of global properties. For example,
the conformal factor and the shift are linear in $r$ by construction
since
$	\psi^{-2}=\frac{r}{R}$ and $\beta^r = \frac{r}{R} \beta$
without any approximation. The leading orders in (\ref{leadord}) arise
from $R\simeq R_0$ and $\beta(R_0) \simeq \frac{2}{R_0} - 1$. Only the
next to leading order terms involve non-linear $r^{1/\gamma}$ terms.

It is somewhat ironic that the Schwarzschild areal radius (which is
not used in 3D numerical simulations) results in a regular Taylor
expansion at $R=R_0$, while the transformation to isotropic
coordinates relabels points such that instead of $r$ the natural
variable is $r^{1/\gamma}$. This happens because of the logarithmic
nature of the combined 1+log and isotropy conditions. The numerics
does not even use isotropic coordinates. Are there globally regular
coordinates that remain linear in radius near the puncture?

Part of the motivation to analyze the local properties for $r=0$ in
spherical symmetry is to prepare a similar study in axisymmetry. For
axisymmetric black holes we do not have an analytic, stationary 1+log
solution, but we can write down a Taylor expansion. Axisymmetric
moving puncture data include the two important cases of a spinning
black hole and of a black hole with linear momentum.  Based on
numerical experiments, the results for single non-moving punctures
generalize to spinning and/or moving punctures, but we do not know yet
how this is reflected in a stationary 1+log slice discussion.

The present paper can also be the starting point for the construction
of initial data for multiple black holes in the moving puncture gauge
since some simplifications and additional details about the analytic
single black solution have been obtained. Standard methods superimpose
single black hole solutions to obtain an ansatz for the solution of
the constraints. In \cite{BauEtiLiu08}, the conformal thin-sandwich
method is considered and the punctures are removed by excision at the
apparent horizon. It is demonstrated that stationary 1+log slices for
binaries constructed with helical Killing vector boundary are not
asymptotically flat (if the data are non-axisymmetric). However, it
may still be possible to construct asymptotically flat initial data
starting from single moving punctures using a different method.

\newpage

\newcommand{\JE}{J.$\,$E.\ }

\begin{acknowledgments}
It is a pleasure to thank M. Hannam, D. Hilditch, S.
Husa, N. \'O~Murchadha, and F. Ohme for discussions.
This work was supported in part by grant SFB/Transregio~7
``Gravitational Wave Astronomy'' of the Deutsche Forschungsgemeinschaft.
\\
\\
This paper is a contribution to the J\"urgen Ehlers memorial volume. 
I have known \JE for a number of years, in particular during his time
as founding director of the Albert Einstein Institute in Potsdam.
\JE was the mentor of my habilitation thesis in 1996, and I
am deeply thankful for many insightful discussions. 
\JE combined great breadth and physical intuition with sharp analytical
thought. His example inspired me to look beyond the numerical methods
and results of numerical relativity to the analytic foundations.  For
example, while at the AEI, S. Brandt and I introduced ``puncture
initial data'' for the numerical construction of general multiple black hole
spacetimes~\cite{BraBru97}. 
While the puncture construction starts with an analytic trick of the
sort that numerical relativists may devise, it is fair to say that the
keen interest in analytical relativity created by \JE at the AEI
induced us to push our analysis one step further. As a result
\cite{BraBru97} connects to \cite{Can79} for an existence and
uniqueness proof for such black hole initial data, using weighted
Sobolev spaces (see also~\cite{BeiOMu94,BeiOMu96,DaiFri01}).
The present work and its
predecessors~\cite{HanHusPol06,Bro07a,GarGunHil07,HanHusOhm08}
represent an example where numerical experiments led to the discovery
of an analytic solution for the 1+log gauge for the Schwarzschild
solution, and the present result, although modest, is of the type
which I believe \JE would have appreciated.
\end{acknowledgments}


\bibliography{refs}

\end{document}